\documentclass[preprint,showpacs,preprintnumbers,amsmath,amssymb]{revtex4}
\usepackage{epsfig}
\usepackage{graphicx}
\usepackage{dcolumn}
\usepackage{bm}

\def\beq{\begin{equation}}
\def\eeq#1{\label{#1}\end{equation}}
\newcommand{\bea}{\begin{eqnarray}}
\newcommand{\eea}{\end{eqnarray}}

\def\eeqn{\end{equation}}
\newcommand\iden{\leavevmode\hbox{\small1\normalsize\kern-.33em1}}
\newcommand{\ptmiss}{p_T\!\!\!\!\!\! /\,\,}

\let\jnfont=\rm
\def\NPB#1,{{\jnfont Nucl.\ Phys.\ B }{\bf #1},}
\def\PLB#1,{{\jnfont Phys.\ Lett.\ B }{\bf #1},}
\def\EPJC#1,{{\jnfont Eur.\ Phys.\ Jour.\ C }{\bf #1},}
\def\PRD#1,{{\jnfont Phys.\ Rev.\ D }{\bf #1},}
\def\PRL#1,{{\jnfont Phys.\ Rev.\ Lett.\ }{\bf #1},}
\def\MPLA#1,{{\jnfont Mod.\ Phys.\ Lett.\ A }{\bf #1},}
\def\JPG#1,{{\jnfont J.\ Phys.\ G }{\bf #1},}
\def\CTP#1,{{\jnfont Commun.\ Theor.\ Phys.\ }{\bf #1},}
\def\JHEP#1,{{\jnfont JHEP \ }{\bf #1},}
\def\NPPS#1,{{\jnfont Nucl.\ Phys.\ Proc.\ Suppl.\ }{\bf #1},}

\begin{document}


\title{Associated production of the Higgs boson and a single top quark in the littlest Higgs model at Large Hadron Collier}

\author{Shuo Yang }
\affiliation{Institute of Theoretical Physics, Chinese Academy of
Sciences, Beijing 100190, China } \email{shuoyang@itp.ac.cn}

\begin{abstract}

In the context of the littlest Higgs model, we study the associated
production of the Higgs boson and a top quark ($th$ production) at
the CERN Large Hadron Collider (LHC). The cross sections for
s-channel, t-channel processes and the relative correction for the
total cross section are presented. In a part of parameter space ,
the cross sections can be distinctly deviated from the predictions
of Standard Model. We also investigate the signal and backgrounds
for the $th$ production at the LHC. However, due to the large QCD
backgrounds, it is not very hopeful to directly observe the $th$
signal in most of the parameter space of LH model. It is found that
with 30 fb$^{-1}$ of integrated luminosity, in a narrow range of the
parameter space $c=0.8$ and $f\leq 1.62$ TeV, a statistical
significance of 3$\sigma$ can be achieved. More than 30 fb$^{-1}$
luminosity will enhance the significance and the possibility of
detecting the signal.
\end{abstract}

\pacs{  }

\maketitle

\section{Introduction}

For the Standard Model (SM) Higgs boson, the main discovery modes at
hadron colliders are the associated production of $Wh$ or $Zh$
\cite{WZH}, the vector boson fusion processes \cite{bosonfusion},
the gluon-gluon fusion mode \cite{gluonfusion}, and the Higgs
associated production with heavy top quark pairs \cite{ttH} or
bottom quark pairs \cite{bbH}. Compared with these modes, the cross
section for the production of the Higgs boson associated with a
single top quark is small \cite{tH,interference,tH2}. ($th$
production Feynman diagrams are shown in Fig.1-Fig.3 .) However,
measurements of the total rates for $Wh$ and $t\bar{t}h$ processes
only test the Higgs coupling to $W$ and the Yukawa coupling to top.
The $th$ production contains the important relative phase
information of the couplings of the Higgs to $W$ and to top, and
$th$ production is sensitive to some new physics
\cite{interference,tH2}. As a class candidate of physics beyond SM,
little Higgs models
\cite{LHidea,MiniMoose,LHM,SU3,LHT,LHreview,LHreview2} predict new
bosons, fermions and scalars. These new particles will contribute
considerably to Higgs production. The Higgs phenomena in little
Higgs models have been extensively studied in the literature
\cite{higgspheno,higgsLH}. In this paper, we study the $th$
production at the LHC in the frame of the littlest Higgs model.

The rest of this paper is organized as follows. In Sec. II we
briefly describe the general features of little Higgs models and
then focus on the littlest Higgs model. The Feynman rules and
formulas relevant to our computation are also listed in this
section. In Sec. III, we study $th$ production in the littlest Higgs
model at the LHC. Finally, we give our conclusions in Sec. IV.

\section{Littlest Higgs Model}
In this section we firstly introduce the general features of little
Higgs models and then describe the littlest Higgs model focusing on
particle content and the couplings relevant to our computation.

 The major motivation of little Higgs theory
\cite{LHidea} is to cure the hierarchy problem of SM. Little Higgs
models suppose that Higgs boson is a Nambu-Goldstone boson (NGB) of
some approximate global symmetry and avoid one-loop quadratically
divergent through collective breaking mechanism \cite{LHidea}. The
collective symmetry breaking means that when one interaction breaks
some of the global symmetries, it still exits unbroken global
symmetry ensuring the Higgs's identity as an exact NGB. The
symmetries protecting Higgs is explicitly broken only when two or
more couplings in the lagrangian are present. In this way, the
collective symmetry breaking protects the Higgs mass from receiving
quadratically divergent radiative corrections at one-loop. The
Coleman-Weinberg potential generates the Higgs potential and
triggers electroweak symmetry breaking (EWSB). The implementation of
collective symmetry breaking results in the prediction of new gauge
bosons, new fermions and new scalars at TeV scale. These new
particles will be produced at the LHC and contribute derivations to
SM processes.

As a realistic implementation of little Higgs idea, the littlest
Higgs (LH) model \cite{LHM} base on a $SU(5)/SO(5)$ coset. The
$SU(5)$ global symmetry is spontaneously broken down to $SO(5)$
global symmetry via a vacuum expectation value (VEV) of order $f$,
simultaneously, and the gauge group $[SU(2)\times U(1)]_1 \times
[SU(2)\times U(1)]_2 $ of $SU(5)$ is broken down to its diagonal
subgroup. The breaks result in new heavy bosons $W_H^{\pm}$, $Z_H$,
$B_{H}$ and new scalar triplet $\Phi$, which acquire masses of order
$f$. In order to implement the collective symmetry breaking
mechanism in top sector, LH model also introduces a pair vector-like
quarks. After EWSB, the physical states in top sector are the top
quark $t$ and the heavy quark $T$. We list the Feynman rules and
formulas \cite{LHhan,smokinghan} relevant to our computation as
below where the terms of order $(\frac{v}{f})^2$ has been neglected:

\bea
 \ \ \
{V}_{W_H^{+\mu}\bar{u}d}=-i\frac{g}{\sqrt{2}}\gamma_{\mu}V_{ud}\frac{c}{s}
P_L \  \ \ \ \ \ \ \ \ \ \ \ \
 {V}_{W_H^{+\mu}\bar{t}b}=-i\frac{g}{\sqrt{2}}\gamma_{\mu}V_{tb}\frac{c}{s} P_L
\eea

\bea
 {V}_{W_H^{+\mu}\bar{T}b}=-i\frac{g}{\sqrt{2}}\gamma_{\mu}V_{tb}\frac{c}{s} \frac{v}{f}\frac{x_{\lambda}^2}{x_{\lambda}^2+1}
 P_L
\  \ \ \ \ \ \ \
 {V}_{W\bar{T}b}=i\frac{g}{\sqrt{2}}\frac{v}{f}\frac{x_{\lambda}^2}{x_{\lambda}^2+1}\gamma_{\mu}V_{tb}
 P_L
\eea

\bea
 \ \ \ \ \ \ \ \ \ \ \ \  \
 {V}_{W_H^{+\mu}W_H^{-\mu}h}=-\frac{i}{2}g^2 v g_{\mu\nu}
\  \ \ \ \ \ \ \ \ \ \ \ \ \ \ \
 {V}_{W_H^{+\mu}W^{-\mu}h}=-\frac{i}{2}g^2 \frac{(c^2-s^2)}{2sc}v g_{\mu\nu}
\eea

\bea
 {V}_{Ht\bar{T}}=-ix_{\lambda}\frac{m_t}{v}P_L
\eea

\bea
 M_{W_{H}}=\frac{gf}{2sc}
\ \ \ \ \ \ \ \ \ \ \
\Gamma_{W_{H}}=\frac{\alpha_e}{96sin\theta_{W}^2}[\frac{96c^2}{s^2}+\frac{cos\theta_{W}^2(c^2-s^2)^2}{s^2-c^2}]
\eea

\bea M_T=(x_{\lambda}+x_{\lambda}^{-1})\frac{m_t}{v}f
 \  \ \ \ \ \ \ \ \ \  \ \ \  \ \ \  \ \  \  \ \
\Gamma_T=\frac{1}{8\pi}\frac{\lambda_1^2}{\lambda_2^2}(\frac{m_t}{\nu})^2M_T.
\eea

In these formulas, $c$ ( $s=\sqrt{1-c^2}$ ) is the mixing parameter
between the $SU(2)_1$ and $SU(2)_2$ gauge bosons, $f$ is the new
symmetry breaking scale, and $x_{\lambda}$=$\lambda_1$/$\lambda_2$,
where $\lambda_1$ and $\lambda_2$ are Yukawa coupling parameters in
top sector. The couplings of $\Phi$ to fermions are suppressed by
$1/f$ and the contribution of $\Phi$ to $th$ production can be
safely ignored, so we do not list the coupling of $\Phi$ here.
\vspace{-0.5cm}
\begin{figure}[htbp]
 \hspace{-3.6cm}
 \psfig{file=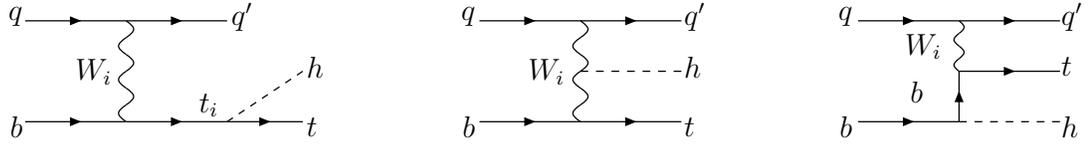,width=20cm}
\caption{Feynman diagrams for $th$ production in t-channel process.}
\end{figure}

\begin{figure}[htbp]
\begin{center}
 \epsfig{file=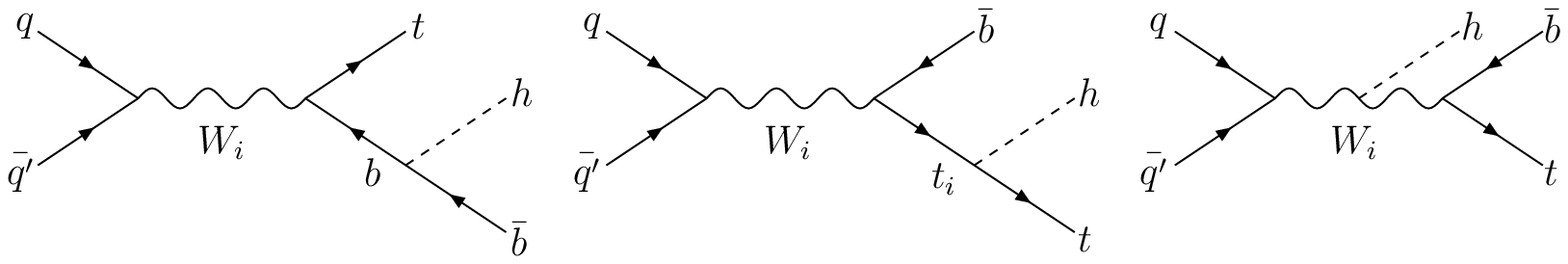,width=18cm}
 \caption{Feynman diagrams for $th$ production in
s-channel process.}
 \end{center}
\end{figure}

\begin{figure}[htbp]
\begin{center}
\vspace{-0.5cm}
 \psfig{file=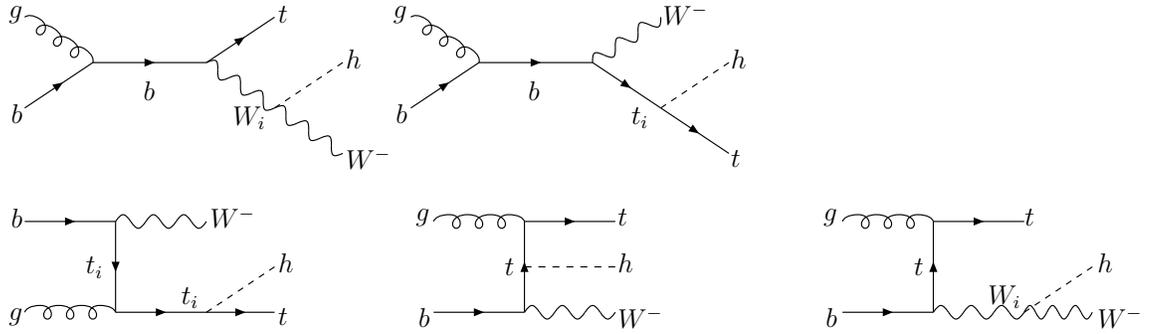,width=18cm}
 \caption{Feynman diagrams for $th$ production in
W-associated process.}
\end{center}
\end{figure}
\section{$th$ Production in the littlest Higgs model at the LHC}

As shown as in Fig.1 - Fig.3, the Higgs in association with a top
quark can be produced at the LHC in t-channel, s-channel and
W-associated process. In these Feynman diagrams, $t_i$ represents
$t$ for SM and represents $t$ and $T$ for LH model, $W_i$ represents
$W$ for SM and represents $W$ and $W_H$ for LH model. In SM, the
t-channel process has the largest cross section and the s-channel
has the smallest cross section. However, in LH model both the cross
sections in s-channel and t-channel can dominate over the
W-associated process in a part of the parameter space as shown
below.

\begin{figure}[htbp]
\begin{center}
 \epsfig{file=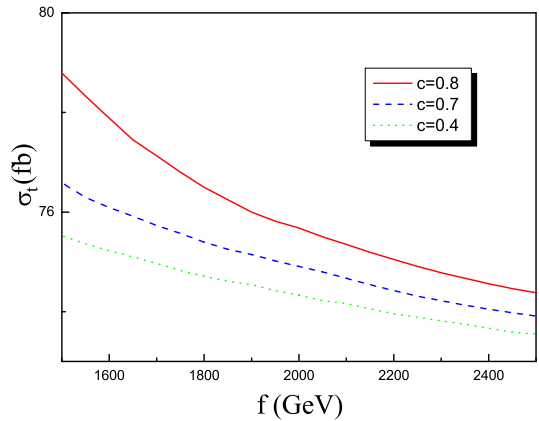,width=8cm}
\end{center}
\vspace{-1cm}
 \caption{Cross section for t-channel $th$ production
as a function of the parameter $f$ for c=0.4 (dot line), c=0.7 (dash
line), c=0.8 (solid line).}
\end{figure}

\begin{figure}[htbp]
\begin{center}
 \epsfig{file=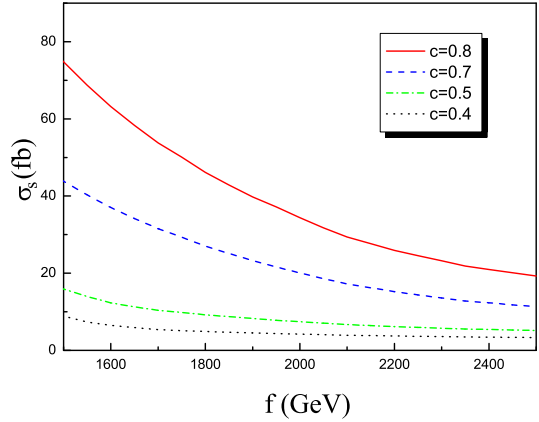,width=8cm}
\end{center}
\vspace{-1cm} \caption{Cross section for s-channel $th$ production
as a function of the parameter $f$ for c=0.4 (dot line), c=0.5 (dash
dot line), c=0.7 (dash line), c=0.8 (solid line).}
\end{figure}
In our computation (We compute $th$ production including both top
quark and antitop production.), we have used Madgraph
\cite{madgraph} package and CTEQ6L \cite{cteq6} parton distribution
functions with the factorization scale $Q=m_h$. The SM input
parameters relevant to our computation are taken as $m_t=174.2$ GeV
\cite{topmass}, $m_h=120$ GeV, $m_Z=91.1876$ GeV,
$sin^2\theta_W=0.2315$, $\alpha_{e}(M_Z) = 1/128.8$ and
$\alpha_s(M_Z) = 0.1176$ \cite{pdg}.

\begin{figure}[h]
\begin{center}
 \epsfig{file=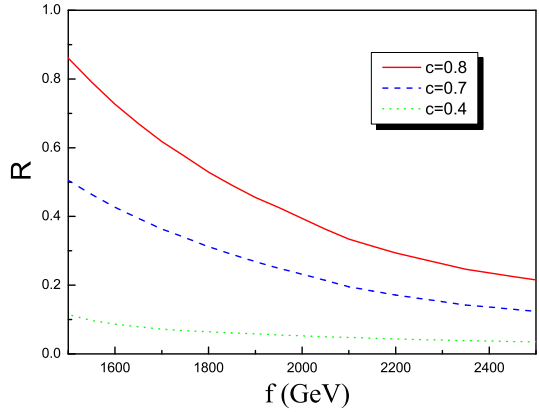,width=8cm}
\end{center}
 \caption{Relative correction R of total $th$ production as a function of the scale parameter
 $f$ for c=0.4 (dot line), c=0.7 (dash line), c=0.8 (solid line).}
\end{figure}

In LH model, the new interactions will contribute the $th$
production and cause deviation from the cross section of SM.
Considering the constraints of the electroweak precision data on LH
model \cite{LHconstrains}, we assumed $x_{\lambda}=1$, $1.5$ TeV
$\leq f\leq 2.5$ TeV and $0.4\leq c \leq0.8$. A low $f$ will result
in a light $B_H$ which is disfavored by $Z'$ limit from Tevatron and
precision electroweak constrains. This can be resolved by
considering alternative embeddings of the two U(1) generators or
only gauging a U(1) group, i.e. the gauge group is $SU(2)_1\times
SU(2)_2 \times U(1)_Y$ \cite{LHconstrains}. Fig.4 and Fig.5 show the
cross sections as a function of the scale parameter $f$ for
different value of mixing parameter $c$ for t-channel and s-channel
$th$ production, respectively. The computation results are very
sensitive to the value of $c$, which because the coupling of $W_H$
to fermions are in proportion to $\frac{c}{s}$ and the cross section
for relevant diagrams will be sensitive to $(\frac{c}{s})^4$. In LH
model, the cross section both in s-channel process and t-channel
process can reach the level of 70 fb in a part of parameter space.
The cross section for W-associated channel is not sensitive to the
variation of parameters and in most of the parameter space the
deviation from the SM prediction is small, so we don't show the
cross section for W-associated $th$ production. Different from the
case in SM where the t-channel process has the largest cross section
and s-channel has the smallest one, in LH model both the cross
sections in s-channel and t-channel can dominate over the
W-associated process in a part of the parameter space. This is
mainly because the couplings of $W_H Wh$ and $Tth$ are suppressed by
mixing parameters as shown in Feynman rules, so the contributes from
new particles are very weak in W-associated channel. In t-channel
and s-channel, the new contributes mainly come from $W_Hqq'$ and
$W_H W_H h$ which are not suppressed. However, the effects from new
particles become weak and the cross sections of $th$ production
converge to the values of SM as $f$ increase and c decrease. In
order to describe the deviation of total $th$ cross section from the
SM prediction, we define a relative correction function
 \bea
 R = \frac{\sigma_{LH} - \sigma _{SM}}{\sigma_{SM}}
\eea and show the $R$ as a function of $f$ for $c=0.4$, $c=0.7$ and
$c=0.8$ in Fig.6. As shown in Fig.6, it is found that in most of the
parameter space, the $th$ cross section in LH model is larger than
the prediction of SM which will enhance the detecting possibility in
this mode.

Now we further consider the signature of $th$ production at the LHC.
In order to provide a efficient lepton trigger the semi-leptonical
decay of top is considered, and the decay of $h \rightarrow
b\bar{b}$ is required for the large branching ratio
\footnotetext[1]{In our calculation, we ignored the corrections to
branching ratio of $h\rightarrow b\bar{b}$ ( $BR(h \rightarrow
b\bar{b}$) ) in LH model. Because in LH model couplings of fermions
and gauge bosons of SM to the Higgs are only shifted by an order of
$(v/f)^2$, and only decay widths of $h\rightarrow \gamma \gamma$, $h
\rightarrow gg$ and $h \rightarrow \gamma Z$ are modified due to new
particles in loops. For a light Higgs, there is no new decay channel
and the dominant decay channel is still $h \rightarrow b\bar{b}$. So
the change of $BR(h \rightarrow b\bar{b})$ is not distinct in most
of the parameter space and the change converges to zero as f
increases. Detailed analysis of Higgs decay in LH model have been
studied in Ref. \cite{higgsLH}.}. In this case for t-channel mode
the signal is $3b+l^{\pm}+\ptmiss \ + $1 forward jet and for
s-channel the signal is $4b+l^{\pm}+\ptmiss$ . However,
ref.\cite{tH2} has pointed that it is difficult to extract the
signal $3b+l^{\pm}+\ptmiss \ + $1 forward jet from the large
backgrounds and the hopeful signal for t-channel is
$4b+l^{\pm}+\ptmiss \ +$1 forward jet where the additional $b$ comes
from the splitting of virtual gluons into $b\bar{b}$ pairs. From a
phenomenological point the difference between s-channel 4b signal
and t-channel 4b signal is that all the jets are high $p_T$ central
jets in s-channel while t-channel signal has a forward jet and a low
$p_T$ $b$-jet. We may separate them by rejecting or tagging a
forward jet. But the s-channel signal will always be accompanied by
a jet coming from QCD radiation. If we consider the s-channel signal
and t-channel signal separately, each of them will be the important
backgrounds for the other. So, in this paper, we consider the signal
is $4b+l^{\pm}+\ptmiss \ +$1jet which include t-channel events and
s-channel events with a radiated jet. We study the signal and
backgrounds in a idealized case that the top is reconstructed with
100\% efficiency, so the $b$ coming from top decay can be separated
from other $b$ quarks.

For the $4b+l^{\pm}+\ptmiss \ +$1jet signal, $tZ\bar{b}j$ with $Z$
decay to $b\bar{b}$ pair, $tb\bar{b}\bar{b}j$ are two of the main
backgrounds. The process $t\bar{t}b\bar{b}\rightarrow
W^+W^-b\bar{b}b\bar{b} \rightarrow jjlvb\bar{b}b\bar{b}$ contributes
to backgrounds when one of the jets from W decay is missed or in the
case that one of the jets from W decay is mistagged as a $b$-jet and
another jet ( including $b$ ) is missed. Due to the large rate of
$t\bar{t}j$, when both $c$ and $s$ quarks from the hadronically
decaying W are mistagged, this process also contributes to the
backgrounds.

Because for many events in our signal the final particles from new
heavy particles' contributions are more energetic and with
high-$p_T$ than the events in SM, So we raise the minimum $p_T^b$ to
20 GeV and loose the forward jet constrain compared with the cuts in
Ref.\cite{tH2}. The cuts are shown as below:

$p_T^b>20$ GeV, $|\eta_b|<2.5$

$p^T_{l,v}>20$ GeV, $|\eta_l|<2.5$

$p^T_j>30$ GeV, $|\eta_j|<5$, $\Delta R_{ij}>0.4$

At least one $b\bar{b}$ pair $|m_{b\bar{b}}-m_h|<22$ GeV.

All $b\bar{b}$ pairs min $m_{b\bar{b}}>$ 90 GeV

 \hspace{-0.7cm} At least one $b\bar{b}$ pair $|m_{b\bar{b}}-m_h|<22$ GeV means that we require at least one
$b\bar{b}$ pair ( not including the $b$ that reconstructs the top )
accord with $|m_{b\bar{b}}-m_h|<22$ GeV. All $b\bar{b}$ pairs min
$m_{b\bar{b}}>$90 GeV means that the invariant mass of all $b$ pairs
( not including the $b$ that reconstructs the top ) is required in
this range.

\begin{table}[h]
\caption{Cross section and events with 30 fb$^{-1}$ of integrated
luminosity for the signal and backgrounds for the $th$ production.
For the signal, the parameter values $c=0.8$, $f=1.6$ TeV,
$x_{\lambda}=1$ are chosen.}

\vspace{0.5cm}
\label{tab:1}       
\begin{tabular}{llllll}
\hline\noalign{\smallskip}
  & Signal & $tZ\bar{b}j$ & $tb\bar{b}\bar{b}$ & $t\bar{t}b\bar{b}$ & $t\bar{t}j$  \\
\noalign{\smallskip}\hline\noalign{\smallskip}
cross section $\sigma$(fb)  $\ $  & 0.482 $\ $& 0.054 $\ $& 0.039 $\ $&0.61$\ $ & 0  \\
events with 30 fb$^{-1}$ $\ $ &14.5$\ $ & 1.6 $\ $& 1.2 $\ $&18.3 $\ $&0 \\
\noalign{\smallskip}\hline
\end{tabular}
\vspace*{-0.5cm}  
\end{table}
\vspace{1.5cm}

At luminosity 30 fb$^{-1}$, only in a very narrow part of the
parameter space $c=0.8$ and $f<1.62$ TeV of LH model, we can achieve
the statistical significance of $3\sigma$. We list the cross section
and events with 30 fb$^{-1}$ which pass the cuts for the signal and
the important backgrounds in table I, in which for the signal the
parameter values $c=0.8$, $f=1.6$TeV, $x_{\lambda}=1$ are taken. The
$b$-tagging efficiency $\epsilon_b=60\%$ , lepton-tagging efficiency
$\epsilon_l=90 \%$ and the mistag probability $\epsilon_c=10\%$,
$\epsilon_s=1\%$ have been included in computation.

\section{Conclusion}
In the frame of the littlest Higgs model we calculate the production
of the Higgs bosons associated a top quark at the LHC. Due to the
new interactions of new gauge boson $W_H$ and new fermion $T$ with
Higgs boson, these new particles can contribute to $th$ production
cross section. We find that in a large part of the parameter space
the cross section of $th$ in s-channel and t-channel production can
deviate largely from the SM prediction. Different from the case in
SM, in LH model both the cross sections in s-channel and t-channel
are larger than that in W-associated process in a part of the
parameter space. We also consider the signal $4b+l^{\pm}+\ptmiss \
+$1jet and backgrounds for the $th$ production. However, due to the
large QCD backgrounds, it is not very hopeful to observe the $th$
signal in most of the parameter space of LH model. It is found that
with 30 fb$^{-1}$ of integrated luminosity, in a narrow range of the
parameter space $c=0.8$ and $f\leq 1.62$ TeV, a statistical
significance of 3$\sigma$ can be achieved. More than 30 fb$^{-1}$
luminosity will enhance the significance and the possibility of
detecting the signal.

 \acknowledgments

The author would like to thank Chun Liu, Jin Min Yang, Chong-Xing
Yue and Lei Wang for useful discussions. This work was supported in
part by the National Science Foundation of China under Grant No.
90503002 and the Project of Knowledge Innovation Program (PKIP) of
Chinese Academy of Sciences, Grant No. KJCX2.YW.W10.

\end{document}